\documentclass[twoside]{article}
\usepackage{fleqn,espcrc2,psfig}

\title{Charmonium in finite temperature lattice QCD}

\author{T. Umeda\address[Hiroshima]{%
     Department of Physics, Hiroshima University,
     1-3-1 Kagamiyama, Higashihiroshima, 739-8526, Japan}%
     \thanks{talk presented by T.Umeda at Lattice 2000,
     Bangalore, India.},
  R. Katayama\addressmark[Hiroshima],
  H. Matsufuru\address{%
     Research Center for Nuclear Physics, Osaka University,
     Mihogaoka 10-1, Ibaraki  567-0047, Japan}
  and
  O. Miyamura\addressmark[Hiroshima]}

\begin{document}

\begin{abstract}
We study hadron properties near the deconfining transition
in the finite temperature lattice QCD.
This paper focus on the heavy quarkonium states, such as
$J/\psi$ meson.
We compare the meson correlators above and below $T_c$
and discuss the possibility of the $c\bar{c}$ bound state
by observing the wave function.
\end{abstract}

\maketitle

\section{Introduction}

It is widely believed that the Quantum Choromodynamics (QCD)
exhibits a phase transition at some temperature $T_c$,
and quarks and gluons confined in the low temperature phase
are liberated to form ``quark gluon plasma''.
At the beginning of 2000, CERN reported that the QGP state
had been created in the heavy ion collision experiment\cite{NA50}.
In this experiment, the $J/\psi$ suppression \cite{MS86} is
regarded as a key signal of QGP formation.
More reliable inspection will be performed in RHIC project at BNL.

On the theoretical side, in spite of various approaches,
we are still far from definite
understanding of charmonium properties near the transition and
the fate of the hadronic states in the plasma phase.
We investigate this problem using lattice QCD, which enables
us to incorporate the nonperturbative effect of QCD.

\section{Our approach}

The difficulties in studying the heavy quarkonium state at finite
temperature can be categorized into two classes:
(i) difficulties to obtain detailed information on the heavy quarkonium
 correlators at finite temperature, and 
(ii) analyses of the obtained correlators.
Although it is difficult to overcome the latter, the former is possible
to be solved by employing the techniques developed in recent years.

At high temperature, sufficient information on the temporal correlator 
requires high resolution in the temporal direction.
To achieve large temporal cutoff with limited computational resources,
we employ the anisotropic lattice, on which the temporal lattice spacing
is finer than the spatial one \cite{Kar82}.
Calculations with a heavy quark need further implementation.
The standard lattice action contains large systematic uncertainty
for the quark mass larger than or comparable to the lattice cutoff.
This problem was reexamined with the light of effective theoretical
interpretation of the Wilson (and clover) quark action by FNAL group
\cite{EKM97}.
Their approach is well-matched with to the anisotropic lattice, since it
naturally introduces the anisotropy in the spatial and the temporal
hopping parameters.
We employ this approach to describe the heavy quark on present lattices,
which have not sufficiently large spatial cutoff compared with the charm
quark mass.

The latter problems, (ii), are much more difficult to be overcome.
This is because one is inevitably enforced to extract the spectral
properties from the data at the short temporal distance, where high
frequency component of the dynamics is significant.
At present, there is no well-established way of lattice simulations
to attack the spectroscopy at $T>0$.
For this reason, lattice studies of hadrons at finite temperature
have been argued on the spatial correlation (screening mass)
\cite{DK87}.
Recently, QCD-TARO Collaboration (including three of us) analyzed the
space-time structure of the mesonic correlators in the temporal
direction to examine the temperature effect on them \cite{taro}.
They caught a sign of the bound state even in the deconfined phase,
as well as an indication of the chiral symmetry restoration.
Their strategy seems one of the best approaches also to the present subject,
and is expected to give us novel information on the fate of charminium
near the phase transition.
Therefore we proceed the analysis with the following steps.

\begin{list}{}{
\setlength{\itemindent}{0.15cm}
\setlength{\leftmargin}{0.2cm}
\setlength{\topsep}{0.15cm}
\setlength{\parsep}{0.0cm}
\setlength{\itemsep}{0.0cm}  }
\item[$\bullet$] At $T=0$ we construct good meson operators which has large 
       overlap with the state of interest.
\item[$\bullet$] Temperature effects are observed in the meson correlator
       between the operators constructed in the previous step.
\item[$\bullet$] We define the ``wave function'' in the Coulomb gauge
       at finite temperature, and observe its $t$-dependence below and above
       $T_c$ to discuss the possibility of the bound state at $T>T_c$.
\end{list}

\section{Fermilab action on anisotropic lattice}

The quark action we use is in the following form.
\begin{equation}
  S_F = \sum_{x,y} \bar{q}(x) K(x,y) q(y),
\end{equation}
\vspace{-0.5cm}
\begin{eqnarray}
& & \hspace*{-0.7cm} K(x,y) = \delta_{x,y}
 \nonumber \\ 
     & &  \hspace{-0.5cm} - \kappa_{\tau} \{ (1-\gamma_4)T_{+4}
                         +(1+\gamma_4)T_{-4} \}
 \nonumber \\
    & & \hspace{-0.5cm}
     - \kappa_{\sigma} {\textstyle \sum_{i}} \{ (r-\gamma_i)T_{+i}
                    +(r+\gamma_i)T_{-i}  \}
 \nonumber \\ 
   & &  \hspace{-0.5cm}
     - ( \kappa_{\sigma} c_E g \sigma_{4i}F_{4i}(x)
               + r \kappa_{\sigma} c_B \frac{1}{2}
                  g \sigma_{ij}F_{ij}(x) ) \delta_{x,y},
\end{eqnarray}
\vspace{-0.4cm}\\
where $T_{\pm \mu} = U_{\pm \mu}(x) \delta_{x\pm\hat{\mu},y}$.
The bare anisotropy  $\gamma_F$ of the fermion field appears
in the ratio of the spatial and the temporal hopping parameters,
$\kappa_{\tau} = \gamma_F \kappa_{\sigma}$.
The Wilson parameter in this form is $r=1/\xi$, where $\xi$ is the
renormalized anisotropy: $\xi=a_{\sigma}/a_{\tau}$.
The hopping parameters are related to the bare quark mass as
$\kappa_{\sigma} = 1/2(m_0 + \gamma_F + 3r)$.
We also define $\kappa$ with
$1/\kappa = 1/\kappa_{\sigma} - 2(\gamma_F+3r-4)$
as the parameter which plays the same role as $\kappa$ on the isotropic
lattice.
As the field strength $F_{\mu\nu}$, we use the standard clover-leaf
definition.
At the tree level, the coefficients of the clover terms, $c_E$ and
$c_B$, are unity.

This action is obtained by generalizing the form in \cite{EKM97}
to the anisotropic lattice, and by setting $r=1/\xi$.
This choice of the Wilson term is adopted so that the action
restores the axial interchange symmetry in the physical unit
as $m_0 \rightarrow 0$.
Therefore we can use the same form in the light
quark region as the $O(a)$ improved Wilson quark action \cite{SW85}.
This is the reason that our form of the action is different from the
form used in \cite{Kla99,Che00}.

On the anisotropic lattice, we need additional tuning of the
parameters called as ``calibration'', to ensure the renormalized
anisotropy is same for the gauge and the fermion fields \cite{Kla99}.
We define the fermionic anisotropy $\xi_F$ using the dispersion
relation of the pseudoscalar and the vector mesons and tune the
value of $\gamma_F$ so that $\xi_F$ coincides with the gauge field
anisotropy $\xi$.

We incorporate the mean-field improvement \cite{LM93}:
$U_i \rightarrow U_i/u_{0\sigma}$ ($i=$1,2,3) and 
$U_4 \rightarrow U_4/u_{0\tau}$ with the mean-field values of the
spatial and the temporal link variables, $u_{0\sigma}$ and $u_{0\tau}$
defined in the Landau gauge.

\section{Result at zero temperature}

Numerical calculation is done on the lattice of sizes $12^3 \times N_t$
where $N_t = 72, 20, 16$ and $12$, with the standard Wilson gauge action
at the coupling $\beta=5.68$ and the bare anisotropy $\gamma_G = 4.0$,
in the quenched approximation \cite{taro}.
On $N_t=72$ lattice, the renormalized anisotropy and the spatial lattice
spacing from the string tension are determined as $\xi = 5.3(1)$ and
$a_{\sigma}^{-1}=0.85(3)$ GeV.
$N_t = 20$, $16$ and $12$ lattices correspond to 
the temperatures  $T \simeq$ 0.93 $T_c$, 1.15 $T_c$ and 1.5 $T_c$,
respectively.
We calculate the meson correlators at $T=0$ with the three sets of the
parameters ($\kappa$, $\gamma_F$) listed in Table~\ref{tab:params}.

\begin{table}[tb]
\begin{center}
\begin{tabular}{ccccc}
\hline\hline
 $\kappa$ & $\gamma_F$ & $m_{Ps}$ [GeV] & $m_V$ [GeV] \\
\hline
 0.0971 & 3.629 & 3.51(16) & 3.56(16) \\
 0.1013 & 3.703 & 3.07(14) & 3.13(14) \\
 0.1055 & 3.765 & 2.67(12) & 2.74(12) \\
\hline\hline
\end{tabular}
\end{center}
\caption{The quark parameters and the meson masses in the physical unit.
The error of the meson masses include the statistical error of
the scale $a_{\tau}^{-1}$.}
\label{tab:params}
\vspace{-0.5cm}
\end{table}

Each value of $\gamma_F$ is determined by the calibration procedure
mentioned above.
Observing the resultant meson masses summarized in Table~\ref{tab:params},
($\kappa$, $\gamma_F$) $=$ (0.1013, 3.703) roughly corresponds to the
charm quark.
In the successive analysis, we use this set of parameters.

To optimize the meson correlator at $T=0$, we employ the
variational analysis.
This is implemented by diagonalizing the correlator matrix
\begin{equation}
 C_{ij}(t) = {\textstyle \sum_{\vec{x}}}
    \langle O_i(\vec{x},t) O_j^\dag(0) \rangle .
\end{equation}
The operator $O_i(\vec{x},t)$ is defined with the smearing 
function $\varphi$ in the Coulomb gauge as
\begin{equation}
 O_i(\vec{x},t)
  = {\textstyle \sum_{\vec{y}}}\, \bar{q}(\vec{x}+\vec{y},t)
    \varphi_i(\vec{y}) \Gamma q(\vec{x},t),
\end{equation}
where the $4\times 4$ matrix $\Gamma$ specifies the quantum number of
the meson.
As the smearing function $\varphi_i$, we use the eigenfunction obtained
by solving the Schr\"odinger equation for the $S$ state,
\begin{eqnarray}
  \left[ -\frac{1}{2m_R}\frac{d^2}{dr^2} +V(r) \right]y(r) &=& E y(r),
  \nonumber \\
 y(r)&=&r\varphi(r),
\end{eqnarray}
where $V(r)$ is the static quark potential obtained on the same lattice,
and $m_R=1.5/2$ [GeV] is used.

\begin{figure}[tb]
\vspace*{0.35cm}
\centerline{\psfig{figure=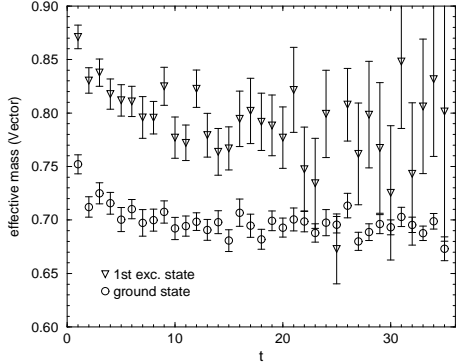,width=6.0cm}}
\vspace{-1.0cm}
\caption{Effective mass plot for the diagonalized correlators
in the vector channel.}
\vspace{-0.4cm}
\label{fig:vanal}
\end{figure}

Figure~\ref{fig:vanal} shows the effective masses of the diagonalized
correlators with two largest eigenvalues (smallest masses).
Both correlators show plateaus beyond $t\sim 10$, though the
statistical fluctuation is large for the correlator corresponding to the
first excited state.
We fit these data at $t=16$--$36$ and find the meson masses in 2$S$ states
as $m_{Ps(2S)}=0.7677(77)$ and  $m_{V(2S)}=0.7794(88)$ in the
lattice unit.
Although these give 1$S$-2$S$ splitting smaller than the
experimental value, considering the large systematic uncertainty on this
coarse lattice, they are not inconsistent results.
We regard the operator corresponds to the ground state in this analysis
as the optimized meson operator at $T=0$.

\section{Result at finite temperature}

\begin{figure}[tb]
\centerline{\psfig{figure=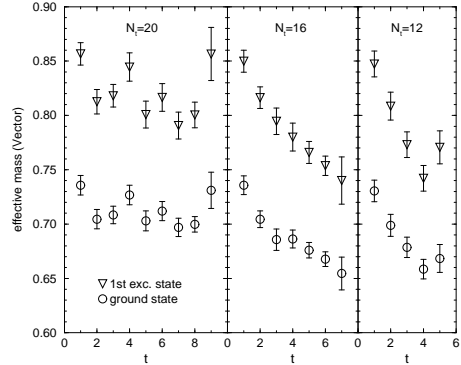,width=6.0cm}}
\vspace{-1.0cm}
\caption{Effective mass plot at finite temperatures
in the vector channel.}
\vspace{-0.4cm}
\label{fig:ep}
\end{figure}

We start the argument at $T>0$ with the temperature
dependence of the correlator between the
optimized operators at $T=0$ difined in the previous section.
Figure~\ref{fig:ep} shows the effective mass plot at finite temperatures
in the vector channel.
Below $T_c$ ($N_t=20$), the plateaus of the effective mass seem
to appear.
Although the effective mass of the ground state is slightly larger than
that at $T=0$, it is difficult to identify the plateau precisely and
determine the mass quantitatively with present statistics.
More detailed analysis with higher statistics may open a stage
to discuss the potential mass shift of charmonium near to the $T_c$
\cite{Has86}.

Above $T_c$ ($N_t=16$ and $12$), the correlators do not show 
any clear plateau in the effective mass plot.
They decrease more rapidly than expected from the behavior 
at $N_t=20$ in both of the correlators correspond to the
ground and the first excited states at $T=0$.
The Ps channel shows similar feature, although the decrease
of the effective mass is slightly milder.
These behavior at least signal significant change in the nature of
the correlators when the system crosses $T_c$.
In \cite{taro}, the (effective) masses increase as $T$ in the
pseudoscalar and the vector channels.
The observed behavior in present work, however, shows qualitatively
different nature of the correlators.

To discuss the spatial correlation between quark and antiquark
at $T>0$, we observe the ``wave function'' normalized at the spatial
origin,
\begin{equation}
 \phi(\vec{r},t) =
  w_\Gamma(\vec{r},t) / w_\Gamma(\vec{0},t) ,
\end{equation}
\begin{equation}
 w_\Gamma (\vec{r},t) = {\textstyle \sum_{\vec{x}}} \langle
   \bar{q}(\vec{x}+\vec{r},t)\Gamma
   q(\vec{x},t)O^\dagger(0)\rangle.
\end{equation}
If there is no bound state, like as with the free quark
propagators, this wave function $\phi(\vec{r},t)$ broadens
with $t$ for any source smearing function.
In contrast, the existence of the stable shape of $\phi(\vec{r},t)$
give us a hint on the existence of the bound state.

\begin{figure}[tb]
\vspace{0.38cm}
\center{\psfig{figure=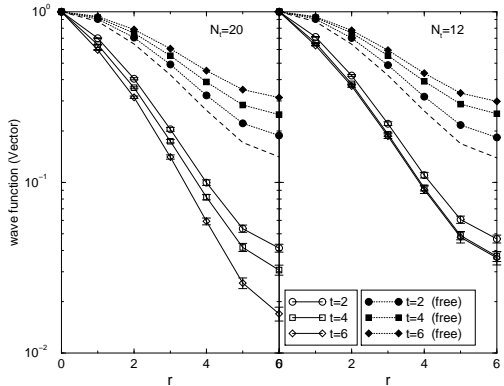,width=6.6cm}}
\vspace{-1.0cm}
\caption{
The $t$-dependence of the wave function normalized at the spatial origin
in the vector channel at $N_t=$20 and 12.
For comparison, the same quantity composed of the free quark propagators
is also shown.}
\vspace{-0.4cm}
\label{fig:wavefunc}
\end{figure}

Figure~\ref{fig:wavefunc} shows the $t$-dependence of the wave function
normalized at the spatial origin in the vector channel at $N_t=$20 and 12.
The source function of these correlators are taken to be wider
than the optimized one,
to show clearly that $\phi(\vec{r},t)$ keeps narrow shape.
At the both $N_t$, the wave function gradually decrease as $t$,
in contrast to the case with the free quark poropagators shown together.
This indicates that the quark and antiquark tend to stay together
even in the deconfined phase, up to 1.5 $T_c$.

These two observations are interesting, and at the same time puzzling.
Obviously, the phase transition causes significant change in the
charmonium correlators, especially in the vector channel.
On the other hand, the result of the wave function suggests 
the persistence of the bound states.
One natural picture is that the mesonic spectral function
still have a peak with rather large width above $T_c$.
This explains both of our observations.
Direct extraction of the spectral function from the lattice data
\cite{spectr} may give us further information.
For the definite understanding of the fate of chamonium at the
phase transition, we need more systematic studies and development
of procedures.

\bigskip

\noindent
The simulation has been done on Intel Paragon XP/S and NEC HSP at INSAM,
Hiroshima University.
This work is supported by the Grant-in-Aide for Scientific
Research by Monbusho, Japan (No.10640272,No.11440080)

\end{document}